# Optical properties of popular dielectric substrate materials in a wide spectral range from far-infrared to ultraviolet

Minjae Kim, Hong Gu Lee, and Jungseek Hwang[*]

*Department of Physics, Sungkyunkwan University, Suwon 16419, Republic of Korea*

AUTHOR INFORMATION

**Corresponding Author** *E-mail: jungseek@skku.edu

## Abstract

We investigated the optical properties of 13 different dielectric materials (slide glass, quartz, $Al_2O_3$ (c-cut), $DyScO_3$ (110), $KTaO_3$ (001), $LaAlO_3$ (001), $(LaAlO_3)_{0.3}$-$(Sr_2AlTaO_6)_{0.7}$ (001) (LSAT), $MgF_2$ (100), MgO (100), SiC, $SrTiO_3$ (001), $TbScO_3$ (110), and $TiO_2$). The single-bounce reflectance spectra of the bulk samples were measured using Fourier transform infrared (FTIR) and monochromatic spectrometers across a wide spectral range, from far infrared to ultraviolet (80–50,000 $cm^{-1}$). Using the Kramers–Kronig analysis, we obtained the optical conductivity and dielectric function of the dielectric materials from their measured reflectance spectra. Moreover, we measured the transmittance spectra of the materials to obtain their bandgaps. We fitted the measured reflectance spectra using the Lorentz model to obtain phononic structures. Each dielectric material exhibits unique phononic structures and optical bandgaps, associated with the composition and crystal structure of the material. The observed optical properties of these dielectric materials provide valuable information for the optical analysis of thin films grown on them.

# 1. Introduction

The optical properties of materials can be determined based on their interaction with light using an optical spectroscopy technique. The optical properties reflect the energetic (electronic and phononic) structure of the electrons in the material; therefore, they are important for designing and developing optical and electronic devices and their applications. In optical spectroscopy, the typically measured optical quantities are the reflectance and transmittance spectra of a material. The reflected and transmitted lights appear after interaction with electrons in the target material through the optical transition process. The optical transitions are transitions of electrons from the filled states (or bands) below the Fermi level to the empty states above it in the material, which are closely related to its phononic and electronic band structures. Particularly, the joint density of states is directly involved in the optical transitions. The optical transitions of a given material are intimately associated with its optical properties (or optical constants). The optical constants are the optical (or frequency-dependent) dielectric constant, optical conductivity, optical index of refraction, absorption coefficient, energy loss function, and so on. The optical constants carry the information on the phononic and electronic band structures. Therefore, the optical properties of a material can be determined by analyzing its measured reflectance and/or transmittance spectra. Several analysis methods for measured reflectance and transmittance spectra use the propagation of light through a material; Maxwell's equations can be used to describe light propagation in matter [1,2]. Many dielectric materials have been used as substrates for growing thin films. There are several well-known thin film growth techniques, such as pulsed laser deposition, chemical vapor deposition, sputtering, and evaporation [3-9]. To study the optical properties of a thin film on a substrate, a suitable substrate and an appropriate method of analysis should be used [10-17]; thus, the optical properties of the substrate materials should be known for selecting a suitable substrate.

In this study, we used a broadband optical spectroscopy technique to investigate 13 dielectric substrate materials widely used for growing thin films: slide glass, quartz, $Al_2O_3$ (c-cut), $DyScO_3$ (110), $KTaO_3$ (001), $LaAlO_3$ (001), $(LaAlO_3)_{0.3}$-$(Sr_2AlTaO_6)_{0.7}$ (001) (LSAT), $MgF_2$ (100), MgO (100), SiC, $SrTiO_3$ (001), $TbScO_3$

(110), and TiO$_2$. We measured the single-bounce reflectance of all samples. The optical constants, including the optical conductivity, were obtained from the measured reflectance using the Kramers–Kronig analysis. We obtained the phononic and electronic structures of the dielectric samples from the measured reflectance and transmittance spectra. For example, the phononic features were obtained by fitting the phonon absorption peaks of the optical conductivities, and the energy bandgaps of the dielectric materials were obtained from measured transmittance spectra. The obtained optical properties will potentially provide useful information to researchers for choosing an appropriate substrate for their thin film growth and studying their applications.

## 2. Experiment

### 2.1 Samples and optical measurements

A total of 13 dielectric samples were purchased from commercial companies. All of the samples, except slide glass, quartz, and SiC, were polished on one side; the polished surfaces were atomically smooth and flat. Each sample size was approximately 5 × 5 mm$^2$. The commercial vacuum-type Fourier transform infrared (Vertex 80v, Bruker) and monochromatic spectrometers (Lambda 950, Perkin-Elmer) were used to cover a wide spectral range, from far-infrared to ultraviolet (80–50,000 cm$^{-1}$ or 10–6.2 eV). The single-bounce reflectance spectra were measured over a wide spectral range at 300 K. For the reflectance measurements, smooth and flat gold and aluminum mirrors were used as references for far-, mid-, and near-infrared (80–13,000 cm$^{-1}$) and visible and ultraviolet (8,000–50,000 cm$^{-1}$) spectral ranges, respectively. These metallic materials have high, roughly constant, and well-known reflectance spectra over a wide spectral range. To obtain the single-bounce reflectance spectrum, we roughened one of the two surfaces of the double-side polished sample to scatter the penetrated light and prevent multiple reflections from the two parallel surfaces of the sample. To create a rough surface, the surface was ground with rough sandpaper or resurfaced with index-matched scotch tape with a rough surface by attaching it to the sample surface. For example, soda lime glass and scotch tape have a similar refractive index ($n \cong 1.5$). To obtain the absolute reflectance ($R(\omega)$) of a sample, the absolute reflectance ($R_{\text{ref}}(\omega)$) of the reference mirror was multiplied by the reflectance with respect to the reference mirror, i.e., $R(\omega) = \frac{P_{\text{sam}}(\omega)}{P_{\text{ref}}(\omega)} \times R_{\text{ref}}(\omega)$, where $P_{\text{sam}}(\omega)$ and $P_{\text{ref}}(\omega)$ are the measured power spectra of the sample and the reference, respectively. Moreover, the transmittance spectra were measured in the visible and ultraviolet regions (12,500–50,000 cm$^{-1}$ or ~1.5–6.2 eV) to accurately determine the energy bandgaps.

**2.2 Data analysis**

Measuring single-bounce reflectance is critical to the analysis of reflectance spectra because, during the usual Kramers-Kronig analysis, the Fresnel equation does not include multiple reflections between the front and back surfaces of the sample. The phase ($\phi(\omega)$) of the reflection coefficient ($\tilde{r}(\omega) \equiv \sqrt{R(\omega)}e^{i\phi(\omega)}$) is related to the amplitude of the reflection coefficient ($\sqrt{R(\omega)}$), where $R(\omega)$ is the reflectance, through the Kramers–Kronig relation as [1,2]

$$\phi(\omega) = \frac{2\omega}{\pi} \int_0^\infty \frac{\ln\sqrt{R(\omega')}}{\omega'^2 - \omega^2} d\omega'.$$

However, the measured reflectance, $R(\omega)$, is in a finite spectral range ($[\omega_L, \omega_H]$) due to the experimental limitations, where $\omega_L$ and $\omega_H$ are the lowest and highest measured frequencies. To perform the Kramers–Kronig integration, the measured reflectance in a finite spectral range must be extrapolated to both zero and infinity. For the extrapolation from $\omega_L$ to 0, we used the Lorentz model fit in a low-frequency region. The extrapolations are shown as dashed lines in Fig. 1(a) and Fig. 2. For the extrapolation from $\omega_H$ to infinity, $R(\omega) \propto \omega^{-1}$ from $\omega_H$ to $10^6$ cm$^{-1}$, and above $10^6$ cm$^{-1}$, the free electron behavior ($R(\omega) \propto \omega^{-4}$) was assumed. The complex index of refraction ($\widetilde{N}(\omega)$) was obtained from the reflection coefficient ($\tilde{r}(\omega)$) using the relationship between $\widetilde{N}(\omega)$ and $\tilde{r}(\omega)$, i.e. $\widetilde{N}(\omega) = \frac{1 - \tilde{r}(\omega)}{1 + \tilde{r}(\omega)}$ for the normal incidence, known as the Fresnel equation. The complex dielectric function ($\tilde{\varepsilon}(\omega)$) was obtained from the index of refraction using the relationship between $\tilde{\varepsilon}(\omega)$ and $\widetilde{N}(\omega)$, i.e., $\tilde{\varepsilon}(\omega) = [\widetilde{N}(\omega)]^2$. The optical conductivity ($\tilde{\sigma}(\omega)$) was obtained from the dielectric function using the relationship between $\tilde{\sigma}(\omega)$ and $\tilde{\varepsilon}(\omega)$, i.e., $\tilde{\sigma}(\omega) = i\frac{\omega}{4\pi}[\varepsilon_H - \tilde{\varepsilon}(\omega)]$), where $\varepsilon_H$ is the high-frequency background dielectric constant. The sharp dropping edge in the measured transmittance spectrum in the high-frequency region was used to determine the bandgap energy. We determined the bandgap energy from the energy where the extrapolation line of the sharp drop in slope met the frequency axis (see Fig. 4(b) and Fig. 6).

## 3. Results and discussions

### 3.1 Single-bounce reflectance, optical conductivity, and phononic structures

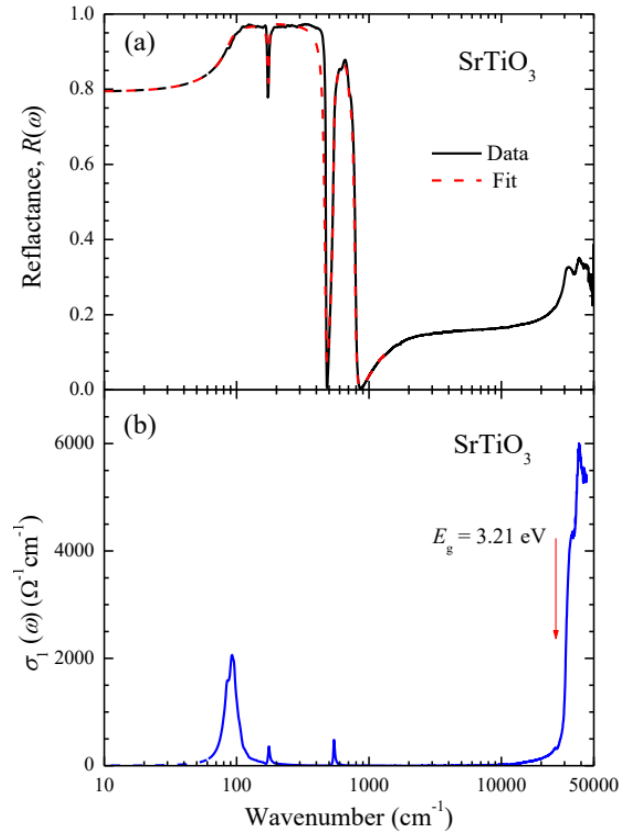

Figure 1. (a) Measured single-bounce reflectance spectrum of SrTiO$_3$ and a Lorentz model fit; herein, we fitted the data up to ~1,500 cm$^{-1}$ to obtain the Lorentz fitting parameters (center frequency, amplitude, and width) of the phonon peaks and (b) the corresponding optical conductivity spectrum.

Fig. 1 shows the measured single-bounce reflectance and optical conductivity spectra of the SrTiO$_3$ sample. The reflectance spectrum shows some characteristic features of dielectric materials: the reflectance ($R(0)$) at zero frequency is always <1.0; some phonon features appear in the low-frequency (far and midinfrared) region. Moreover, there exists a flat region above the phonon features up to the bandgap energy; a sharp upturn appears near the bandgap energy. The region of the phonon features depends on the masses of the elements of the dielectric material; generally, heavier elements produce lower phonon frequencies. The phonon features can be clearly seen in the optical conductivity; they appear as sharp peaks in the infrared region below the bandgap energy. For SrTiO$_3$, three major phonon peaks can be observed (Fig. 1(b)). The reflectance spectrum is fitted using the Lorentz model to obtain a precise phononic structure and to extrapolate the reflectance from the lowest measured frequency ($\omega_L$) to zero frequency. The complex dielectric function is described using the Lorentz model as [1,2]

$$\tilde{\varepsilon}(\omega) = \sum_i \frac{\Omega_{p,i}^2}{\omega^2 - \omega_i^2 + i\omega\gamma_i} + \varepsilon_H,$$

where $\omega_i$, $\Omega_{p,i}$, and $\gamma_i$ are the center (or resonance) frequency, plasma frequency (or amplitude), and width of the *i*-th Lorentz mode, respectively; $\varepsilon_H$ is the high-frequency background dielectric constant. Note that the square of the plasma frequency is proportional to the electron density, which is involved in the Lorentz mode. The single-bounce reflectance ($R(\omega)$) at normal incidence is written as $R(\omega) = \left|\frac{\sqrt{\tilde{\varepsilon}(\omega)}-1}{\sqrt{\tilde{\varepsilon}(\omega)}+1}\right|^2$. During the fitting of the reflectance spectrum, the peak positions in the optical conductivity were used as a guide; we fit both the reflectance and optical conductivity of each sample using the same fitting parameters. The complex optical conductivity is written in terms of the complex dielectric function as $\tilde{\sigma}(\omega) = i\frac{\omega}{4\pi}[\varepsilon_H - \tilde{\varepsilon}(\omega)]$. The optical conductivity was obtained from the measured single-bounce reflectance spectrum of SrTiO$_3$ using the Kramers–Kronig analysis. The red vertical arrow indicates the bandgap energy, which was directly determined from the measured transmittance spectrum of SrTiO$_3$ (Fig. 4(b)).

The measured single-bounce reflectance and optical conductivity spectra of the remaining 12 samples are shown in Figs. 2 and 3, respectively. As was mentioned previously, the overall shapes of the reflectance are similar for all samples; the phonon features are below ~1,500 cm$^{-1}$ and a flat region between the phonon features and the upturn, which was associated with the bandgap. The fitted reflectance (red dashed line) for each sample below ~1,500 cm$^{-1}$ was also shown in Fig. 2. We used several Lorentz (or phonon) modes to fit the reflectance, as previously discussed. Tables 1 and 2 display the obtained fitting parameters; $\varepsilon_H$ is the high-frequency background dielectric constant, and the other parameters are the center frequency, plasma frequency (or amplitude), and width of each Lorentz mode. Fig. 3 shows the real parts of the optical conductivity spectra of the remaining 12 samples. The vertical red arrow shows the bandgap energy of each sample, which was determined from the measured transmittance spectrum (see Fig. 4(b) and Fig. 6). As we can see in the figure, it may be difficult to estimate the correct bandgap energy from the optical conductivity. It is worth noting that because the slide glass and quartz are made from the same compound, SiO$_2$, they exhibit similar phonon mode frequencies. However, phonons in quartz are sharper than those in slide glass, and additionally, the bandgap energy of quartz was larger than that of slide glass (Tables 1 and 3) because glass is amorphous, whereas quartz is a single crystal.

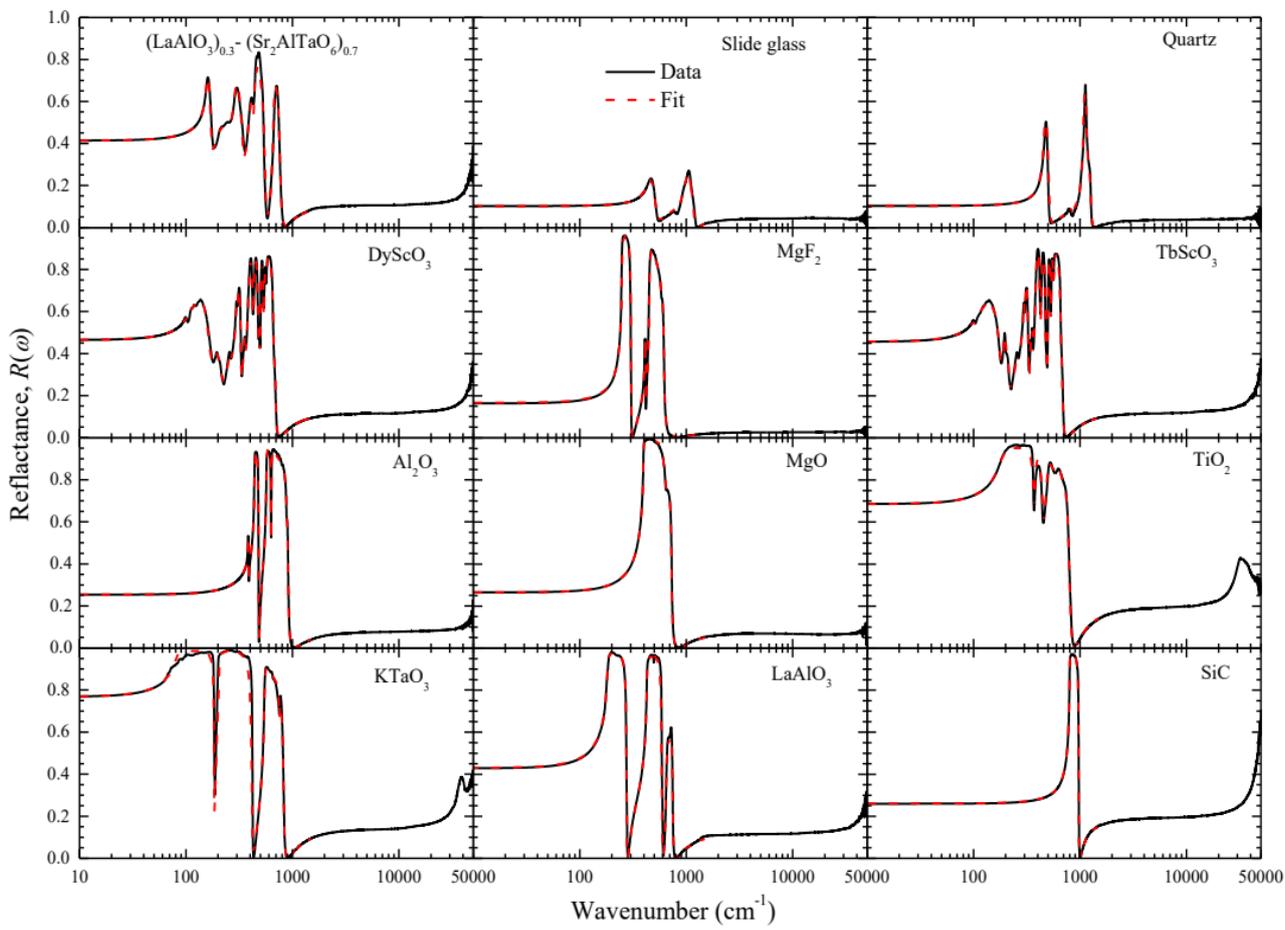

Figure 2. Measured single-bounce reflectance spectra of the remaining 12 samples and their Lorentz model fits (red dashed lines) in the low-frequency region below 1,500 cm$^{-1}$.

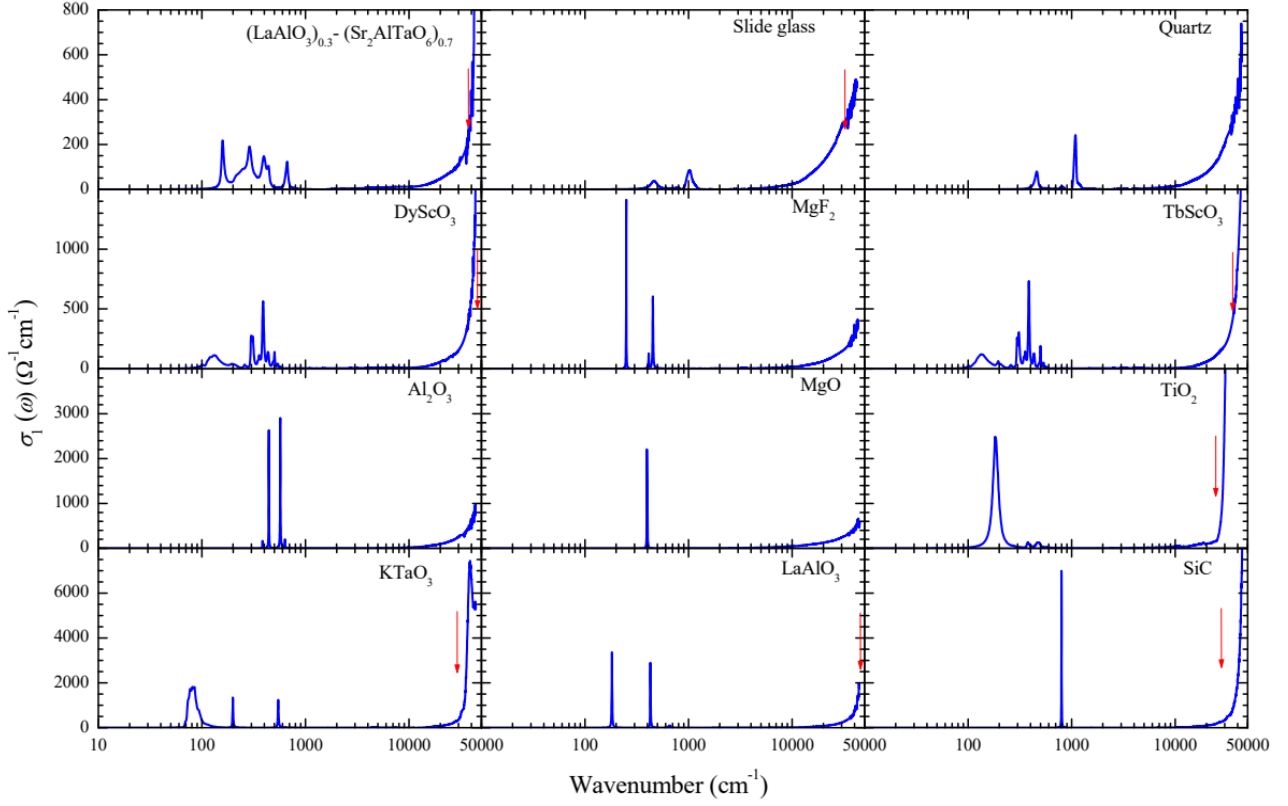

Figure 3. Real part of the optical conductivity spectra of the remaining 12 samples; vertical red arrows indicate the bandgap energies obtained from the measured transmittance spectra.

Table 1. Phonon modes: Lorentz model fitting parameters of 6 samples.

| | Samples | | Slide glass | | | Quartz | | | Al$_2$O$_3$ | | | DyScO$_3$ | | | KTaO$_3$ | | | LaAlO$_3$ | | |
|---|---|---|---|---|---|---|---|---|---|---|---|---|---|---|---|---|---|---|---|---|
| | $\varepsilon_H$ | | 2.35 | | | 2.20 | | | 3.10 | | | 4.20 | | | 4.75 | | | 4.15 | | |
| $\omega_i$ | $\Omega_{p,i}$ | $\gamma_i$ | 466.2 | 418.6 | 77.0 | 459.3 | 429.3 | 28.5 | 384.7 | 156.9 | 2.5 | 101.0 | 66.9 | 4.0 | 80.8 | 1,182.5 | 5.3 | 182.0 | 694.5 | 2.3 |
| | | | 775.7 | 104.9 | 42.0 | 577.0 | 129.0 | 120.0 | 442.8 | 719.4 | 3.3 | 113.0 | 80.0 | 5.0 | 205.4 | 596.4 | 2.5 | 428.4 | 861.0 | 4.0 |
| | | | 983.3 | 420.3 | 80.9 | 807.0 | 160.0 | 42.0 | 571.3 | 975.6 | 5.5 | 119.8 | 111.2 | 7.0 | 546.0 | 813.7 | 9.5 | 494.4 | 50.0 | 8.1 |
| | | | 1,030.0 | 604.3 | 94.7 | 1,078.0 | 816.4 | 32.2 | 633.9 | 244.4 | 5.3 | 132.0 | 464.8 | 35.0 | 576.5 | 385.1 | 153.5 | 651.8 | 350.0 | 21.4 |
| | | | 1,112.5 | 301.3 | 114.9 | 1,188.5 | 306.7 | 111.0 | 653.2 | 286.7 | 208.1 | 197.6 | 256.4 | 36.0 | 754.1 | 60.8 | 18.7 | 696.2 | 113.0 | 36.7 |
| | | | | | | | | | 887.7 | 17.3 | 12.4 | 261.6 | 126.9 | 10.0 | | | | 673.0 | 121.8 | 30.0 |
| | | | | | | | | | | | | 300.0 | 280.8 | 7.0 | | | | 216.4 | 56.0 | 20.6 |
| | | | | | | | | | | | | 309.0 | 416.5 | 13.0 | | | | | | |
| | | | | | | | | | | | | 347.9 | 134.5 | 12.0 | | | | | | |
| | | | | | | | | | | | | 357.0 | 249.9 | 13.3 | | | | | | |
| | | | | | | | | | | | | 391.0 | 669.1 | 12.4 | | | | | | |
| | | | | | | | | | | | | 433.5 | 352.8 | 16.6 | | | | | | |
| | | | | | | | | | | | | 486.0 | 188.7 | 18.4 | | | | | | |
| | | | | | | | | | | | | 502.0 | 266.0 | 10.0 | | | | | | |
| | | | | | | | | | | | | 541.5 | 180.5 | 16.4 | | | | | | |
| | | | | | | | | | | | | 569.9 | 100.0 | 16.4 | | | | | | |
| | | | | | | | | | | | | 673.2 | 77.4 | 43.7 | | | | | | |

Table 2. Phonon modes: Lorentz model fitting parameters of remaining 7 samples.

| LSAT | | | MgF$_2$ | | | MgO | | | SiC | | | SrTiO$_3$ | | | TbScO$_3$ | | | TiO$_2$ | | |
|---|---|---|---|---|---|---|---|---|---|---|---|---|---|---|---|---|---|---|---|---|
| 3.90 | | | 1.95 | | | 3.00 | | | 6.45 | | | 5.45 | | | 4.20 | | | 6.70 | | |
| 158.0 | 338.7 | 9.7 | 250.0 | 372.1 | 1.5 | 397.3 | 1,030.0 | 1.7 | 798.0 | 1,400.0 | 4.5 | 90.0 | 1,532.0 | 17.3 | 101.5 | 72.0 | 6.0 | 182.2 | 1,858.0 | 27.1 |
| 215.4 | 192.2 | 27.8 | 412.0 | 203.3 | 5.9 | 646.4 | 132.7 | 99.1 | | | | 175.7 | 345.2 | 5.3 | 124.0 | 265.0 | 18.0 | 377.3 | 381.5 | 17.5 |
| 245.3 | 75.8 | 12.7 | 452.0 | 483.2 | 6.8 | | | | | | | 544.9 | 598.9 | 13.0 | 136.0 | 240.0 | 16.0 | 472.6 | 550.2 | 41.3 |
| 251.3 | 578.7 | 94.7 | 509.2 | 144.3 | 126.5 | | | | | | | 557.6 | 389.3 | 36.6 | 150.0 | 354.0 | 40.0 | 582.3 | 211.4 | 74.9 |
| 288.4 | 514.1 | 28.2 | 584.1 | 19.4 | 14.4 | | | | | | | 618.0 | 130.9 | 50.2 | 195.0 | 170.0 | 12.0 | 684.5 | 111.8 | 80.6 |
| 397.8 | 601.4 | 47.7 | 656.4 | 33.1 | 40.2 | | | | | | | 714.0 | 47.8 | 33.2 | 208.0 | 130.0 | 17.0 | 813.7 | 104.9 | 80.4 |
| 438.4 | 289.1 | 17.3 | 716.7 | 107.5 | 72.4 | | | | | | | 950.0 | 237.4 | 300.0 | 258.0 | 100.0 | 7.0 | | | |
| 462.1 | 70.0 | 9.6 | | | | | | | | | | | | | 297.0 | 248.4 | 5.0 | | | |
| 659.2 | 503.6 | 27.9 | | | | | | | | | | | | | 308.0 | 458.3 | 12.0 | | | |
| 784.0 | 73.6 | 72.5 | | | | | | | | | | | | | 345.0 | 175.0 | 9.2 | | | |
| | | | | | | | | | | | | | | | 355.0 | 250.0 | 10.0 | | | |
| | | | | | | | | | | | | | | | 386.0 | 646.6 | 10.0 | | | |
| | | | | | | | | | | | | | | | 432.0 | 310.7 | 13.8 | | | |
| | | | | | | | | | | | | | | | 484.0 | 165.2 | 19.3 | | | |
| | | | | | | | | | | | | | | | 500.0 | 297.9 | 8.9 | | | |
| | | | | | | | | | | | | | | | 535.0 | 155.2 | 10.0 | | | |
| | | | | | | | | | | | | | | | 564.2 | 100.0 | 16.4 | | | |
| | | | | | | | | | | | | | | | 669.4 | 69.5 | 50.4 | | | |

## 3.2 Dielectric functions, measured transmittance spectra, and bandgaps

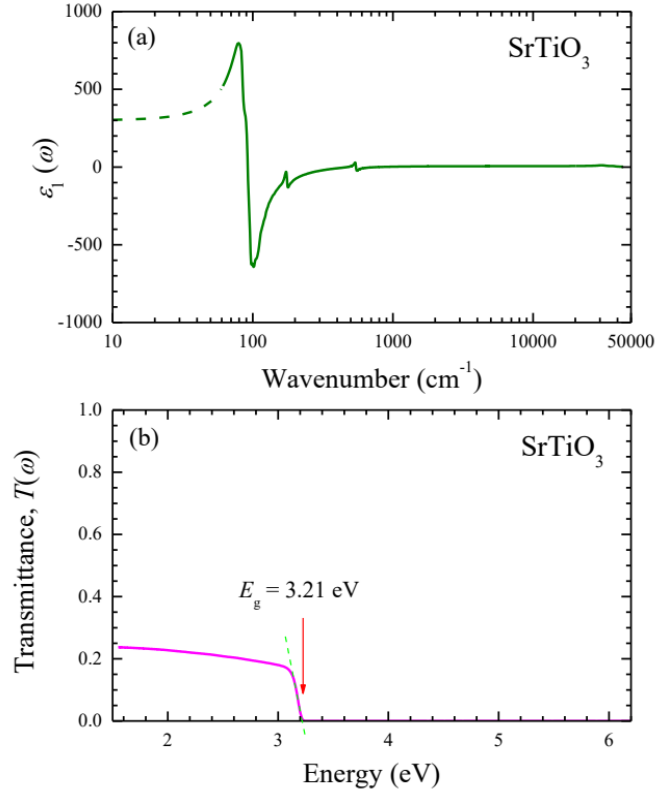

Figure 4. Dielectric function and measured transmittance spectra of SrTiO$_3$; the green dashed line is the linear extrapolation of the steep dropping edge near $T(\omega) = 0$; the dashed line in panel (a) is the portion extrapolated to a frequency of zero.

Fig. 4 shows the dielectric function and measured transmittance spectra of the SrTiO$_3$ sample. In Fig. 4(a), the dielectric function shows three major phonon features near the positions of the phonon peaks in the optical conductivity (see Fig. 1(b)). As the frequency decreases from the high to low energy regions, the overall dielectric function increases because each (phonon) absorption mode produces an additional background dielectric constant. Therefore, the static dielectric constant ($\varepsilon_0$), the dielectric function at zero frequency, is supposed to be the largest. However, near the resonance phonon frequency, the dielectric function exhibits a dramatic change, including maximum frequency dependence at the resonance frequency. The static dielectric constant was estimated from the extrapolation of the dielectric function to a zero frequency. Because free charge carriers do not exist in dielectric material, the dielectric function approaches a positive value at a near-zero frequency. The dielectric function of SrTiO$_3$ exhibits a typical frequency-dependent behavior of dielectric materials. Its static dielectric constant (~300) is one of the largest among dielectrics because there is a low-frequency phonon mode located at ~90 cm$^{-1}$. A large static dielectric constant is important for applications such as capacitors. In Fig. 4(b), the transmittance spectrum of SrTiO$_3$ measured over a wide spectral range from ~1.5 to 6.2 eV is shown. We determined the bandgap energy ($E_g$) from this spectrum, as is shown in the figure. We linearly extrapolated the slope of the sharp drop in transmittance near $T(\omega) = 0$, and we determined the bandgap from the energy value at $T(\omega) = 0$, as shown in the figure (the green dashed line). Note that features like the tail next to the extrapolation line on the high energy side were ignored when the bandgap energy was determined.

Fig. 5 shows the real parts of the dielectric functions of the 12 remaining samples. The static dielectric constant was closely associated with the properties (the center frequency, amplitude, and width) of the lowest-lying phonon mode. In general, a low center frequency gives a higher static dielectric constant. Table 3 shows a summary of the obtained static dielectric constants for all 13 samples along with the reported ones [18-29]. The obtained static dielectric constants agree quite well with the reported ones.

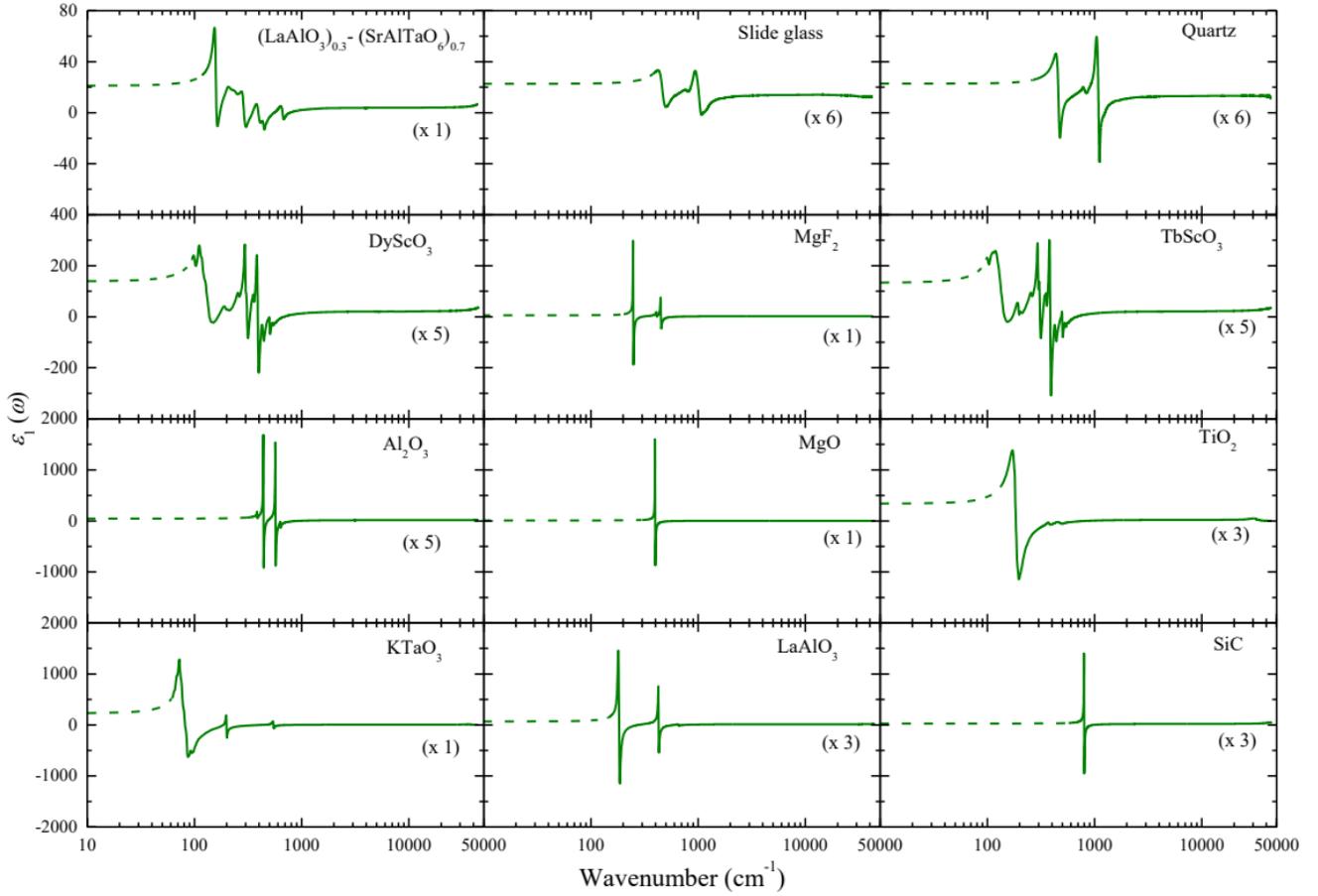

Figure 5. Dielectric functions of the remaining 12 samples. The number with × stands for the magnification of the raw dielectric function to better display it on the same vertical scale for the samples in the same row. The dashed lines are the extrapolations to a zero frequency.

Fig. 6 shows the measured transmittance spectra of the 12 remaining samples. The transmittance spectra of all 13 samples were measured over a wide spectral range from ~1.5 to 6.2 eV. Most samples are one-side polished; only three (slide glass, quartz, and SiC) are double-side polished. The scattering at the rough side of a sample increases with increasing energy, because the Rayleigh scattering law ($I_s(\omega) \propto \omega^4$) results in a decrease in transmittance, where $I_s$ is the scattering intensity. Rayleigh scattering can be observed in the transmittance spectra of the 10 one-side polished samples. As we previously explained, the decreasing slope near $T(\omega) = 0$ in the measured transmittance spectra was used determining the bandgap energy ($E_g$) of the dielectric material, as marked with red vertical arrows in the figure. Due to the strong absorption by the bandgap, no transmission was observed in energies higher than the bandgap energy. Notably, the bandgap of

TbScO$_3$ might have been underestimated because there was a small but nonzero transmittance at energies higher than the bandgap energy (insets of the panels for DyScO$_3$ and TbScO$_3$). The reported bandgaps of TbScO$_3$ are 6.1 eV [30] and 6.5 eV [31], which are larger than the estimated bandgap. Therefore, we recorded the bandgap energy as >4.58 eV in Table 3. Two samples (DyScO$_3$ and TbScO$_3$) with $f$-electrons show many absorption dips in their transmittance spectra in the high-frequency region. The dips of DyScO$_3$ and TcScO$_3$ are the $f$–$f$ transitions of Dy$^{3+}$ [32] and Tb$^{3+}$, respectively. Table 3 shows the determined bandgap energies of all 13 samples along with their reported values [29,30,33-37], showing good agreement.

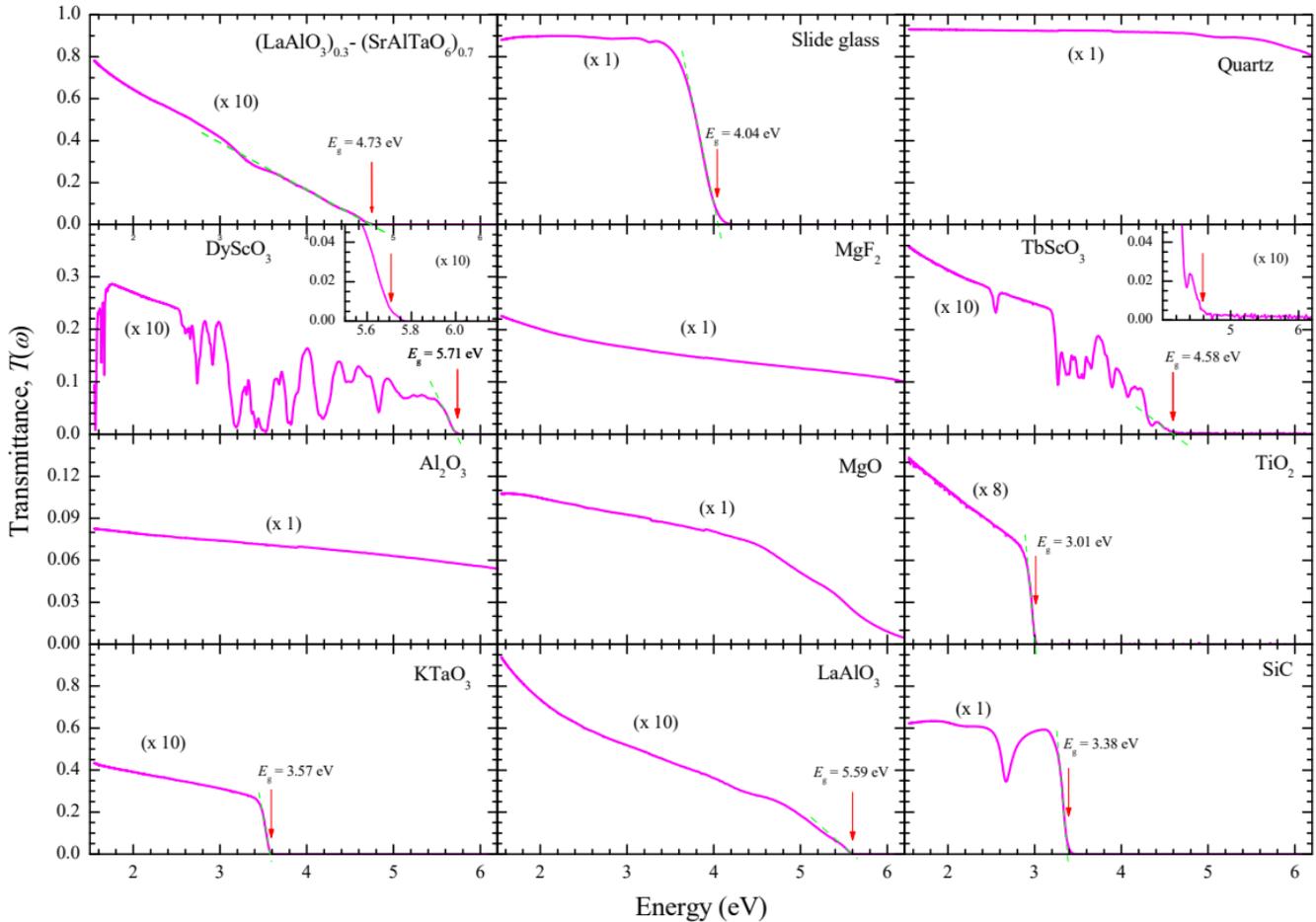

Figure 6. Measured transmittance spectra of the 12 remaining samples; in the figure, the x value stands for the magnification of the measured transmittance to better present it on the same vertical scale; green dashed lines are the extrapolation of the sharply decreasing slope near $T(\omega) = 0$

Table 3. Summary of static dielectric constants ($\varepsilon_0$) and bandgap energies ($E_g$) for all 13 samples

| Samples | Obtained static dielectric constant, $\varepsilon_0$ | Reported static dielectric constant, $\varepsilon_0$ | Obtained bandgap, $E_g$ (eV) | Reported bandgap, $E_g$ (eV) |
|---|---|---|---|---|
| Slide glass | 3.8 | 5~6 [18] | 4.04 | 4.408 [33] |
| Quartz | 3.9 | 3.72~3.90 [19] | Out of range | |
| $Al_2O_3$ | 9.2 | 9 [20] | Out of range | |
| $DyScO_3$ | 27.9 | 22 [21] | 5.71 | 5.7 [34] |
| $KTaO_3$ | 230.3 | 209 [22] | 3.57 | 3.64 [35] |
| $LaAlO_3$ | 22.9 | 24 [23] | 5.59 | 5.7 [34] |
| LSAT | 21.2 | 22 [24] | 4.73 | 4.72 [36] |
| $MgF_2$ | 5.6 | 5.53 [25] | Out of range | |
| MgO | 9.8 | 9.9 [26] | Out of range | |
| SiC | 9.5 | 9.97 [27] | 3.38 | 3.28 [29] |
| $SrTiO_3$ | 300.6 | 300 [28] | 3.21 | 3.2 [37] |
| $TbScO_3$ | 26.7 | 26 [21] | >4.58 | 6.1, 6.5 [30,31] |
| $TiO_2$ | 113.2 | 86 [29] | 3.01 | 3.2 [38] |

## 4. Conclusions

We investigated the optical properties of popular dielectric substrate materials using broadband optical spectroscopy. We obtained the dielectric function and optical conductivity spectra from the measured single-bounce reflectance spectra for 13 dielectric samples: slide glass, quartz, $Al_2O_3$ (c-cut), $DyScO_3$ (110), $KTaO_3$ (001), $LaAlO_3$ (001), $(LaAlO_3)_{0.3}$-$(Sr_2AlTaO_6)_{0.7}$ (001) (LSAT), $MgF_2$ (100), MgO (100), SiC, $SrTiO_3$ (001), $TbScO_3$ (110), and $TiO_2$. Additionally, we obtained the bandgaps from the measured transmittance spectra for all samples. Each dielectric material exhibited its own optical properties, i.e., phononic and electronic structures, static dielectric constant, and bandgap energy. The unique properties of each dielectric material contribute to a deep understanding of the material and provide valuable information for obtaining the optical and electronic properties of thin films grown on the substrate.

**Acknowledgement:**

This research was supported by the National Research Foundation of Korea (NRFK Grant Nos. 2021R1A2C101109811) and BrainLink program funded by the Ministry of Science and ICT through the NRFK (2022H1D3A3A01077468).